\documentclass{jpsj2}

\title{Microwave Absorption of Surface-State Electrons on Liquid $^3$He}

\author{Hanako \textsc{Isshiki}$^{1,2}$\thanks{E-mail address: hisshiki@riken.jp}, 
Denis \textsc{Konstantinov}$^{1}$,
Hikota \textsc{Akimoto}$^{1}$,
Keiya \textsc{Shirahama}$^{2}$,
Kimitoshi \textsc{Kono}$^{1}$,
}

\inst{$^{1}$Low Temperature Physics Laboratory, RIKEN, Hirosawa 2-1, Wako 351-0198, Japan \\
$^{2}$Department of Physics, Keio University, Hiyoshi 3-14-1, Yokohama 223-8522, Japan \\
}

\abst{
We have investigated the 
intersubband transitions of surface state electrons (SSE) on liquid $^3$He induced by microwave radiation at temperatures from 1.1~K down to 0.01~K.
Above 0.4~K, the transition linewidth is proportional to the density of $^3$He vapor atoms. 
This proportionality is explained well by Ando's theory, in which the linewidth is determined by the electron - vapor atom scattering. 
However, the linewidth is larger than the calculation by a factor of 2.1. 
This discrepancy strongly suggests that the theory underestimates the electron - vapor atom scattering rate.
At lower temperatures, the absorption spectrum splits into several peaks.
The multiple peak structure is partly attributed to the spatial inhomogeneity of the static holding electric field perpendicular to the electron sheet.
}

\kword{Two-dimensional electron system, microwave absorption, liquid $^3$He, free surface}

\begin{document}
\maketitle

\section{Introduction} 
\subsection{Surface - State Electrons and Quantum Computing}
Quantum computing has been of great interest as a key technology of the future. 
However, there are a number of hurdles to overcome in order to realize a quantum computer. 
In particular, long decoherence time and scalability are required for the computing elements, qubits. 
Two - dimensional electrons formed on a free surface of liquid helium, so called surface state electrons (SSE), provide a promising candidate for qubits~\cite{Dykman99,Mukharskii,Dahm02}.

On a liquid helium surface, electrons are trapped and form discrete surface states, by an attractive image potential and a strong repulsive surface barrier caused by the Pauli exclusion principle~\cite{2DEonHe97}.
The attractive part of the potential energy for an electron is expressed as $V(z)=-\Lambda e^2/4 \pi \epsilon_0 z$ for $z > 0$. 
Here we denote the coordinate perpendicular to the surface as $z$. 
The coefficient $\Lambda$ is $(\epsilon -1)/4(\epsilon +1)$, $\epsilon_0$ the vacuum permittivity, and $\epsilon$ the relative permittivity of liquid helium. 
At $z \leq 0$, $V(z) \sim 1 {\mathrm {eV}}$, and this large repulsive potential prevents the penetration of electrons into the liquid. 
The Schr\"{o}dinger equation for the vertical motion is identical to the radial wave equation of a hydrogen atom. 
Hence the energy spectrum is easily obtained as $E_n=-R/n^2$ ($n=1,2,\ldots $) with the effective Rydberg energy $R=\Lambda e^2/8 \pi \epsilon_0 a_B$, and the effective Bohr radius $a_B=4 \pi \epsilon_0 \hbar^2/m_e \Lambda e^2$. 
Here $m_e$ is the effective electron mass. 
Two isotopic helium liquids, $^4$He and $^3$He, give different values for $R$ and $a_B$ due to the difference in liquid density. 
$R$ is $-7.6$ and $-4.2$~K, and $a_B$ is 7.6 and 10.1 nm for $^4$He and $^3$He, respectively. 
As the energy difference between the ground and first excited states is much larger than 1 K, the electrons stay in the ground state at modestly low temperatures. 
Since the electrons are free to move parallel to the surface, they constitute a two - dimensional system, and the hydrogenic energy levels become a series of the two - dimensional subbands. 
The helium surface can trap the electrons up to the areal density $n_s \le 2 \times 10^9$ ${\mathrm {cm}}^{-2}$.

Platzman and Dykman~\cite{Dykman99}, and later Lea \textit{et al.}~\cite{Mukharskii} and Dahm \textit{et al.}~\cite{Dahm02} proposed an idea to utilize SSE on liquid helium as a scalable quantum computer. 
Their proposed quantum computer consists of many single electrons trapped independently by an array of micrometer - sized metal electrodes immersed in liquid helium. 
The ground ($n=1$) and first ($n=2$) excited states of an electron are used as the two quantum states necessary for a qubit. 
The operation of a qubit is performed by applying microwave pulses resonant with the $n = 1 \leftrightarrow 2$ intersubband transition. 
The intersubband transition frequency can be tuned finely by static electric field applied perpendicular to the surface. 
The field is applied perpendicular to hold the electrons on the surface, and it results the Stark effect to the hydrogenic states. 
This Stark tuning is conveniently utilized for operating the qubits individually by controlling the electrostatic potentials of each metal electrode. Strong Coulomb repulsion allow qubits to interact. 
Finally, the 
results of the quantum computing are read out by extracting the electrons from the surface~\cite{Dykman99}. 
Lea \textit{et al.} also proposed to use metal electrodes made of single electron transistors (SET)\cite{Mukharskii}. 
In this way the quantum information can be read out without losing the SSE. 

The SSE quantum computer has several important advantages compared to other possible qubit systems. 
In particular, the unique properties of the SSE on helium will result in long decoherence time and excellent scalability. 
Previous in-plane electronic transport measurements~\cite{Shirahama95} revealed that the SSE mobility is the highest known, $\mu \sim 10^8 ~{\mathrm {cm^2/(Vs)}}$. 
This is because the free surface of liquid helium is extremely smooth and flat, and has no impurities. 
Scatterers which determine the electron mean free time in the in-plane transport are helium vapor atoms and capillary surface wave quanta, ripplons. These scattering mechanisms may contribute to the decoherence of the quantum states and cause a serious problem in realistic qubit operations. 
However, the vapor atoms completely vanish at temperatures below 0.7 K and 0.3 K for liquid $^4$He and $^3$He, respectively, and do not contribute to the transport at all~\cite{Iye,Shirahama95}. 
The electron - ripplon scattering will then be the only mechanism which prevails in decoherence, but both the number and energy of thermally excited ripplons decrease at temperatures below 0.1 K.   
Moreover, the SSE quantum computer is highly scalable, because the in-plane separation between the electrons is of the order of $\mu$m. 
The fabrication of the electrodes is feasible with present microfabrication technology.

For the realization of the 
SSE qubits, one must develop the following electron manipulation techniques: (1) Trapping a single electron on micrometer - sized electrodes or SET, (2) Controlling the $n = 1 \leftrightarrow 2$ intersubband transition, and producing Rabi oscillations between the two states, and (3) Detecting the quantum states of individual electrons. 
Papageorgiou \textit{et al.} succeeded in trapping and counting one to several electrons on a suspended $^4$He film by means of a SET~\cite{Papageorgiou05}. 
It is now crucial to study the intersubband transitions. 
We have studied the SSE intersubband transitions on liquid $^3$He induced by microwave radiation at temperatures from 1.1~K down to 0.01~K.
\subsection{Intersubband Transitions}
The intersubband transitions were studied by several groups over the last few decades. 
Grimes {\textit{et al}}. first studied the intersubband transitions on a $^4$He surface~\cite{Grimes74}. 
They observed the intersubband transitions by measuring resonant absorption spectra at several fixed frequencies from 130 to 220~GHz, using the Stark tuning technique. 
The resonant frequencies agreed well with a calculation based on a simple model potential. 
They found that the transition linewidths are determined by electron collisions with $^4$He vapor atoms down to 1.2~K, at which temperature the measurement was limited. 
The intersubband transitions was also observed by Lambert and Richards~\cite{Lambert81} at far-infrared regimes.
Interest in the intersubband transitions has been rekindled by the proposals of quantum computing mentioned above. 
Collin \textit{et al.}~\cite{Collin02} measured the $n=1 \leftrightarrow 2$ transition linewidth down to 0.3~K. 
The linewidth below 0.8~K was found to be determined by electron - ripplon scattering. 
The linewidth taking into account both the vapor atom and ripplon scattering mechanisms was calculated by Ando~\cite{Ando78}, explaining qualitatively the observed temperature dependence of the linewidth. 

All of previous direct absorption studies were performed on liquid $^4$He. 
The study of the intersubband transitions on the liquid $^3$He surface is even more interesting and important. 
Liquid $^3$He has an advantage in cooling the SSE far below 1 mK by collisions of quasiparticles at the free surface and solid walls, while on liquid $^4$He both cooling and measuring the temperature of the SSE is limited to 10 mK. 
Moreover, the large viscosity of liquid $^3$He may strongly attenuate the ripplons which cause decoherence. 
The authors' group has intensively studied the in-plane transport of the SSE (the Wigner crystal) on both normal and superfluid $^3$He, and discovered that it is very sensitive to the scattering of $^3$He quasiparticles at the free surface~\cite{SSEon3He}. 
It is therefore interesting to explore the effect of $^3$He quasiparticles on the intersubband transitions. 
Volodin and Edel'man~\cite{Edelman74} estimated the linewidths from the change in the electron mobility when the SSE were excited to the $n=2$ state by incident microwaves. 
Although the estimated linewidths for both $^3$He and $^4$He were consistent with Ando's theory, their measurement was based on an indirect method and was limited to a narrow temperature range. 
No systematic studies of the intersubband transitions on liquid $^3$He were performed. 

Here we report mainly on the properties of the linewidth of the intersubband transitions on liquid $^3$He. 
This paper is organized as follows. In the next section, we describe the details of the microwave absorption measurement at very low temperatures. In the subsequent section we give data and analyses of the absorption spectra, and discuss the linewidths.  Comparisons with Ando's theory and with the previous results by others are also described.
\section{Experimental}
\subsection{Microwave Cell and Cryogenics}
\begin{figure}[tbp]
\begin{center}
\includegraphics[width=7cm,keepaspectratio]{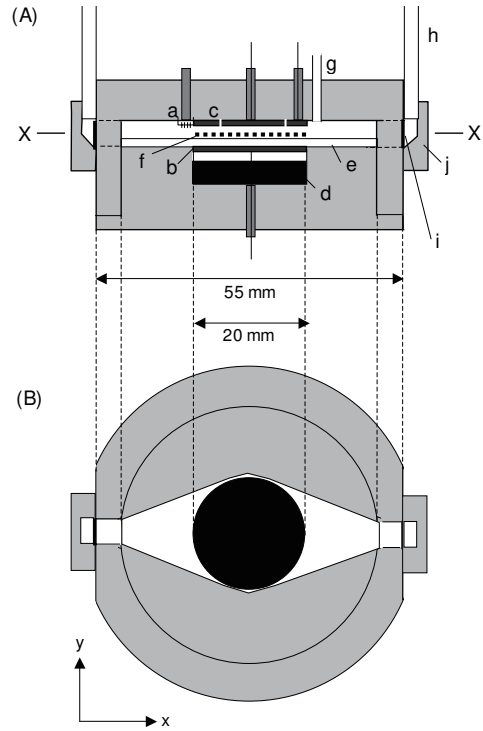}
\end{center}
\caption{\label{SAMPLECELL} A schematic drawing of the sample cell.
(A) A cross sectional view: (a) tungsten filament, (b) bottom electrode, (c) top electrode, (d) silver powder heat exchanger, (e) liquid helium, (f) surface state electrons, (g) filling capillary, (h) plastic waveguide, (i) Kapton sheet, and (j) waveguide holder.
 (B) A top view at the cross section XX' in (A).}
\end{figure}

In order to obtain the absorption spectra at the intersubband resonances, we introduce microwaves to a SSE sample cell mounted on the mixing chamber of a dilution refrigerator, and perform the sensitive detection of the transmitted microwaves.  
In Fig. \ref{SAMPLECELL}, we show a schematic view of the microwave cell.
The body of the cell is made of oxygen-free copper. Two circular electrodes (20 mm in diameter), which are made of circuit boards (thin copper coated on epoxy plate), are located at both the ceiling and bottom of the inner space. The separation between the plates is 3.0 mm. The upper electrode is divided into a disk (14 mm in diameter) and a surrounding ring separated by a gap of 0.1 mm. They constitute a Corbino disk, and are used to measure the in-plane SSE conductivity, which is not discussed in this paper. In this work the Corbino conductivity measurement is used only to check the generation of the SSE and its persistence during the microwave measurement. 
For trapping and Stark-tuning the SSE, a positive bias voltage $V_{\textrm{B}}$ is applied to the lower electrode, while the upper one is grounded.
Liquid $^3$He or $^4$He is introduced into the cell through a capillary, while monitoring the depth of the liquid $d$ by measuring the change of the capacitance between two electrodes with a capacitance bridge.
The height of the free surface is set to 1.5 mm above the lower electrode. 
The liquid depth $d$ is determined with an accuracy of $\pm$0.05~mm. The accuracy is limited mainly by the capacitance measurement.

Electrons are generated by thermionic emission from the tungsten filament. The thermionic emission is performed at a fixed temperature, where helium vapor is dense enough to decelerate the emitted electrons, 0.6 K and 1.2 K on liquid $^3$He and $^4$He, respectively. 
The SSE are formed so that they shield the bias electric field above the surface. 
Therefore, the areal electron density $n_s$ is determined by $V_{\textrm{B}}$ at the time of the electron emission, and is expressed as $n_s=\epsilon_0 \epsilon V_{\textrm{B}}/\mathrm{e}d$.
In our experiment, $n_s$ is varied from $1.9 \times 10^7$ to $1.0\times 10^8$ cm$^{-2}$.
During the microwave measurement, $V_{\textrm{B}}$ is increased from the value at the electron emission, to produce the Stark shift, while $n_s$ is kept unchanged.

The sample cell contains a silver powder heat exchanger to have high thermal conductance between liquid $^3$He and the mixing chamber below 0.1 K.
We measure the temperature of the cell with a carbon resistance thermometer mounted on the outside of the cell. 
\subsection{Electronics and Measurement}
\begin{figure}[tbp]
\begin{center}
\includegraphics[width=7cm,keepaspectratio]{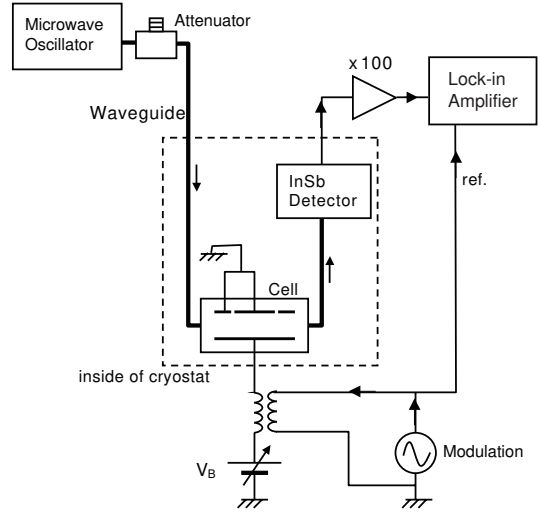}
\end{center}
\caption{\label{MeasurementSystem} A block diagram of the measurement system.}
\end{figure}

Figure \ref{MeasurementSystem} shows a block diagram of the microwave measurement system.
For microwave generation, we have employed a Gunn diode oscillator and a backward wave oscillator (BWO) working at room temperature.
The former produces microwaves at a fixed frequency of 130~GHz, while the frequency of the BWO is varied from 110 to 170~GHz.
The measurement frequencies are 130~GHz and 169~GHz for liquid $^3$He and $^4$He, respectively. 
The short-term stability of the microwave frequency was less than 10~MHz for both oscillators.
The microwave power was controlled over three orders of magnitude (mW $ \sim \mu $W) by a variable attenuator placed after the oscillator.
The microwaves are then introduced from the cryostat top to the sample cell by a rectangular flexible dielectric waveguide~\cite{Keycom} of 5.0 $\times$ 2.5~mm$^2$. The loss in passing the waveguide is less than 10~dB, which is low enough for the present work.
The incident microwave passes the cell through two rectangular windows (5.0 $\times$ 2.5~mm$^2$) sealed with 50~$\mu$m-thick Kapton sheets 
(Fig.~\ref{SAMPLECELL}(i)),
and it is detected by an InSb bolometer mounted at the still of the dilution refrigerator.
The microwave power is measured from the voltage drop across the bolometer under a constant current.
The sensitivity of the bolometer is 10~kV/W, and the response time is about 1 $\mu$s.
The attenuation of microwave power from the oscillator to the bolometer is about 30~dB. 

We have employed the Stark tuning technique to detect the intersubband transitions with a fixed microwave frequency. This is done by sweeping the bias voltage $V_{\textrm{B}}$. Far from the edge of the electron system, the holding electric field $E_{\perp}$  
is given by
\begin{eqnarray}
E_{\perp }&=& E_{\textrm{ext}}+E_{\textrm{image}}  \nonumber  \\
&=& \frac{\epsilon V_{\textrm{B}}}{\epsilon (D-d)+d}
+\frac{n_s e}{2\epsilon_0}\frac{\epsilon (D-d) -d}{\epsilon (D-d)+d},
\label{equationEtotal}
\end{eqnarray}
where $D$ is the distance between the upper and the lower electrode (3.0 mm).
The first term $E_{\mathrm{ext}}$ is the external electric field created by the potential difference between the two electrodes,
while the second term $E_{\mathrm{image}}$ is due to the image charges created by the SSE.
When $d=\epsilon D/ (\epsilon +1)$, $E_{\perp }$ becomes independent of $n_s$, and the transition frequency is determined only by $V_{\textrm{B}}$.
We have attempted to realize this condition by finely adjusting the liquid level and the parallelism between the free surface and the electrodes, but a small deviation from the condition might remain. We will discuss this problem in the subsequent section. 

We have adopted an amplitude modulation technique to obtain the absorption signal by the SSE.
This is done by modulating the holding voltage $V_{\textrm{B}}$ with an audio frequency (10.03~kHz) during the sweep.
The modulated component of the output voltage of the InSb bolometer is detected by a lock-in amplifier.
The voltage at the lock-in amplifier is therefore proportional to the first derivative of the power absorbed by the SSE with respect to $V_{\textrm{B}}$.
In order to measure the resonance linewidth precisely, we set the time constant of the lock-in amplifier to be one tenth of the time required to scan 
from the positive to the negative peak in the derivative of a single absorption resonance.
The modulation amplitude is also set to be one tenth of the separation of the two peaks.
Furthermore, the microwave power is reduced so that the power broadening effect is negligible and the linewidth is independent of the radiation power.
\section{Results}
\subsection{Frequencies of Intersubband Transitions}
\begin{figure}[tbp]
\begin{center}
\includegraphics[width=7cm,keepaspectratio]{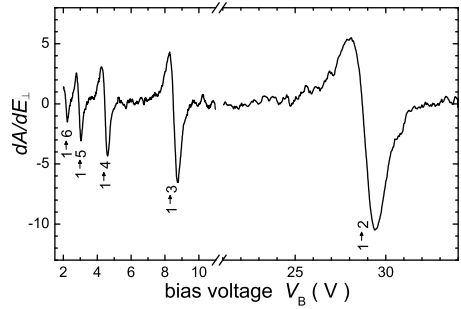}
\end{center}
\caption{\label{HighExcitations} An experimental trace of the derivative of the absorption
signal for the SSE on $^3$He as a function of the bias voltage $V_{\textrm{B}}$ at $T$=0.65~K and $n_{\mathrm{s}} = 1.2 \times 10^7$~cm$^{-2}$.
The microwave frequency is 130~GHz.
For the transition from the ground to the first excited state, the microwave power entering the cryostat $P$
is 93~$\mu $W.
For other transitions below 10~V, $P$
$= 320$~$\mu$W.
The magnitude of the signal is normalized by the modulation amplitude and the microwave power.
}
\end{figure}

We have observed the resonant absorption due to the intersubband transitions from the ground ($n=1$) 
to the five excited states, $2 \le n \le 6$.
Figure~\ref{HighExcitations} shows a typical spectrum of the derivative signal at 0.65~K, as a function of $V_{\textrm{B}}$, i.e. the holding electric field.
The transition frequencies are in excellent ageeement with the calculation of the transition frequencies from the formula of the subband energies, in which the change in the holding electric field is taken into account.
On the $^4$He surfce, the frequencies also agree with the calculation, as in the previous studies by Grimes \textit{et al.}~\cite{Grimes74} and Lambert and Richards~\cite{Lambert81}. 

We have found that the positions of the resonances depend slightly on electron density and temperature. 
The density dependence was investigated for the SSE on $^3$He, and the change is about 1.7~\% of the resonance frequency when $n_s$ varies from $1.9 \times 10^7$ to $1.0 \times 10^8$~cm$^{-2}$.
For the temperature dependence, the variation is about 1.4~\% from 0.01~K to 0.45~K on $^3$He, and is about 6.6~\% from 0.3~K to 1.5~K on $^4$He.
The density dependence indicates that the helium depth $d$ does not satisfy the relationship $d=\epsilon D/ (\epsilon +1)$.
From eq.~(\ref{equationEtotal}), the helium level is estimated to be 0.05~mm above the middle. As this deviation is comparable to the accuracy in the capacitance - based level measurement,
it is difficult to adjust the liquid depth to satisfy the above relationship.
On the other hand, the temperature dependence might be attributed to a variation in the helium level due to the thermal expansion of liquid $^3$He and $^4$He.
However, the estimated level change is an order of magnitude too small to explain the observed temperature dependence. 

\subsection{Linewidths}
The measurement of the transition linewidth is the main subject of the present work. To obtain the linewidth, we have measured the absorption spectrum of the $n = 1 \leftrightarrow 2$ transition at temperatures varied from 1.1 to 0.01 K.
We show typical spectra in Fig.~\ref{LINESHAPE_Tdep_3He}. The lineshape has two features: (1) It is asymmetric. (2) As the temperature is lowered, a single absorption peak splits into several peaks.

\subsubsection{Asymmetry in the Resonances}
At temperatures above 0.5~K, the lineshape is close to a Lorentzian, but an asymmetry is seen as a difference in the positive and negative peak heights in the derivative signal. Such an asymmetry was also observed by Lambert and Richards ~\cite{Lambert81} at far-infrared frequencies. 
The asymmetry is attributed to interference between the radiation directly transmitted through the cell and the radiation which has coherently scattered from the SSE. We have attempted to fit the lineshape with a modified Lorentzian formula containing asymmetry,
\begin{equation}
F(\omega)=A\frac{\left( \omega - \omega_0 -B \gamma \right)^2}{\left(\omega-\omega_0 \right)^2+\gamma^2}.
\label{AsymmetricFormula}
\end{equation}
Here $\omega$ is the angular frequency of excitation. The transition frequency is denoted as $\omega_0$, $A$ is the magnitude of the resonance, and $B$ the coefficient of the term which results in the asymmetry. The parameter $\gamma$ is the full Lorenzian linewidth at half maximum. In the fitting to the $V_{\textrm{B}}$ scan spectra, we can replace $\omega$ with $V_{\textrm{B}}$, since the scan range is so narrow that $V_{\textrm{B}}$ is a linear function of $\omega$. 
The coefficient of the conversion of frequency ($f = \omega/2\pi$) to $V_{\textrm{B}}$ is 1.90 GHz/V, which is determined by a numerical calculation.
The fitted curves are shown in Fig.~\ref{LINESHAPE_Tdep_3He}(a). We have found that at $T > 0.5$~K the resonances are fitted perfectly with eq.~(\ref{AsymmetricFormula}). 
Therefore, we use this formula to obtain the linewidths $\gamma$ from the resonance spectra of all temperatures, although the origin of the asymmetry in the observed resonances is not fully understood. It should be noted that the interference-induced asymmetric resonance is analogous to the Fano resonance seen in various systems\cite{Fano61}. 

\subsubsection{Multiple Peak Structure}
\begin{figure}[tbp]
\begin{center}
\includegraphics[width=7cm,keepaspectratio]{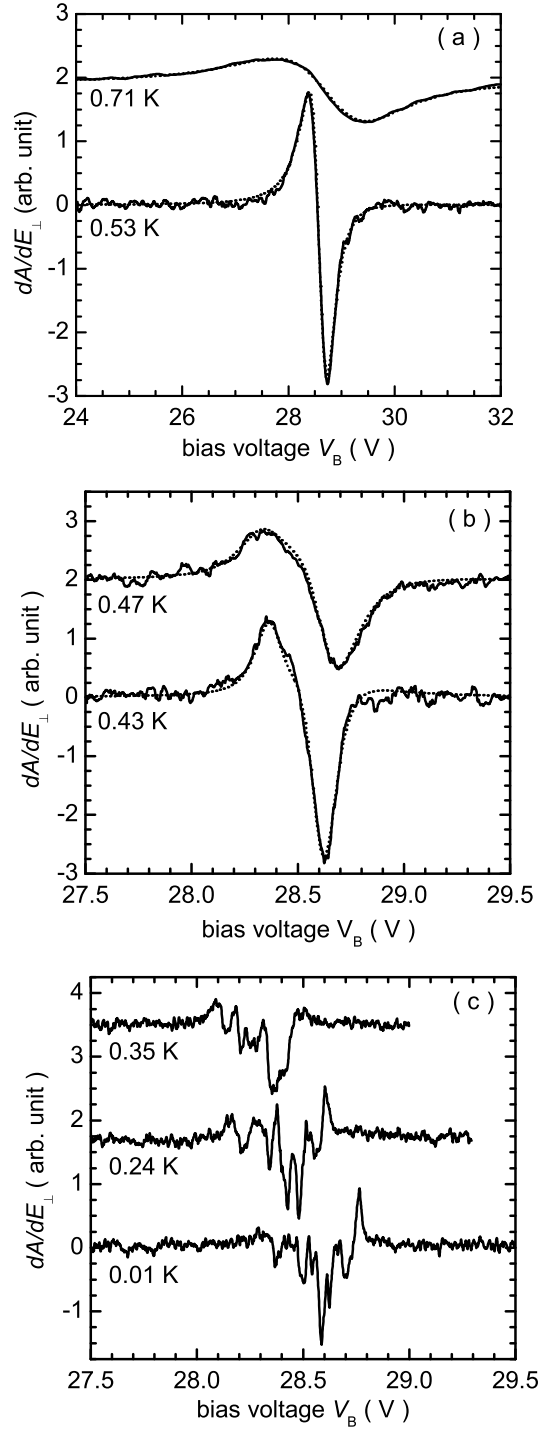}
\end{center}
\caption{\label{LINESHAPE_Tdep_3He}The derivatives of absorption 
spectra on liquid $^3$He at $n_\mathrm{s}=7.0 \times 10^7$~cm$^{-2}$.
(a) Data at 0.71 and 0.53~K, (b) 0.47 and 0.43~K, (c) 0.35, 0.24, and 0.01~K.
The dotted lines in (a) and (b) are the fits to eq. ({AsymmetricFormula}).}
\end{figure}

The absorption linewidth decreases with decreasing temperature.
The peak becomes distorted below 0.5~K as shown in Fig.~\ref{LINESHAPE_Tdep_3He}(b). 
This distorted peak results from a splitting of the peak into two. 
Furthermore, as the temperature is lowered below 0.4~K, the transition signal splits into several peaks. 
This is shown in Fig.~\ref{LINESHAPE_Tdep_3He}(c).

At 0.4~K $<T<$ 0.5~K, the experimental data are fitted well with two asymmetric Lorentzians (eq. (\ref{AsymmetricFormula})).
The linewidth $\gamma$ is obtained by the two peak fitting assuming different resonant frequencies $\omega_0$ and different values of the coefficient $B$.
The results of the two-peak fitting are shown by the dotted lines in Fig.~\ref{LINESHAPE_Tdep_3He}(b). 
The experimental lineshapes are fitted very well with the two peaks in which the linewidths are the same and the resonant frequencies differ by 0.7~\verb|%|.

At $T < 0.4$~K, however, the peaks split into more than three. This makes the fitting and derivation of the linewidth extremely difficult. 
In the present work, we discuss the linewidths obtained by the single peak fitting ($T > 0.5$~K) and the double peak one ($0.4 < T < 0.5$~K).

We have also observed the asymmetric behaviors and the multiple peak structures on the $^4$He surface. 
On liquid $^4$He, the two-peak structure appears at $ 0.7 < T < 1.0$~K. The two asymmetric Lorentzian fittings have also been successful in this temperature regime.
Below 0.7 K, the peaks split into several peaks, and the linewidth is no longer obtained. These behaviors are strikingly similar to the $^3$He case. 

\subsubsection{Temperature Dependence of Linewidth}
\begin{figure}[tbp]
\begin{center}
\includegraphics[width=7cm,keepaspectratio]{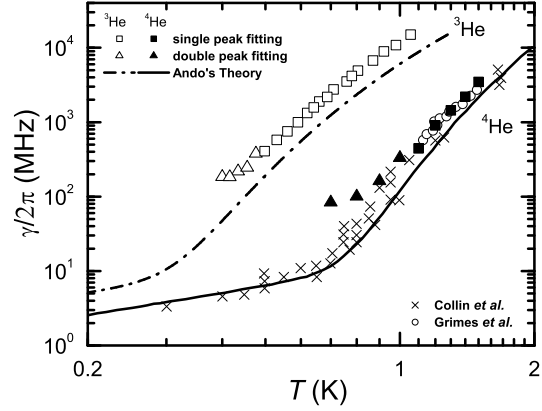}
\end{center}
\caption{\label{LINEWIDTH}
The SSE linewidth $\gamma$ for both liquid $^3$He and $^4$He, as a function of temperature.
The data are taken at $n_s=7.0 \times 10^7$ cm$^{-2}$.
The data obtained by the single asymmetric peak fitting (see text) are expressed as ($\square$) for $^3$He and ($\blacksquare$) for $^4$He, and the ones obtained by the double peak fitting are denoted as ($\triangle$) for $^3$He and ($\blacktriangle$) for $^4$He, respectively.
The experimental data obtained by Grimes \textit{et al.}~\cite{Grimes74} and Collin \textit{et al.}~\cite{Collin02} are also shown.}
\end{figure}

In Fig. \ref{LINEWIDTH}, we show the linewidth $\gamma$ as a function of temperature. Both the data taken on liquid $^3$He and $^4$He are plotted. 
We have found that the linewidth is independent 
of the areal electron density in the range of our measurement (1.9 $\times 10^7 < n_s < 1.0 \times 10^8$~cm$^{-2}$) within the experimental accuracy. 
We discuss the data taken at $n_s=7.0 \times 10^7$ cm$^{-2}$ as a representative.

The linewidth $\gamma$ decreases exponentially with temperature.
Ando~\cite{Ando78} showed theoretically that the linewidth is dominated by the collisions of the SSE with the helium vapor atoms. 
Since the vapor density decreases exponentially with decreasing temperature, $\gamma$ also obeys an exponential law. 
In the same figure, we show lthe inewidths calculated by Ando. 
The temperature dependence of our linewidth data on liquid $^3$He agrees well with the calculated linewidths. 
We conclude that the linewidth is dominated by the electron - vapor atom collisions. 

The linewidth on $^4$He in the present work also agrees with the Ando's calculation. It also agrees with the previous experimental results by Grimes {\textit {et al.}} and Collin {\textit {et al.}}~\cite{Grimes74,Collin02}
above 0.9~K, but a discrepancy is seen below the temperature. 

\begin{figure}[tbp]
\begin{center}
\includegraphics[width=7cm,keepaspectratio]{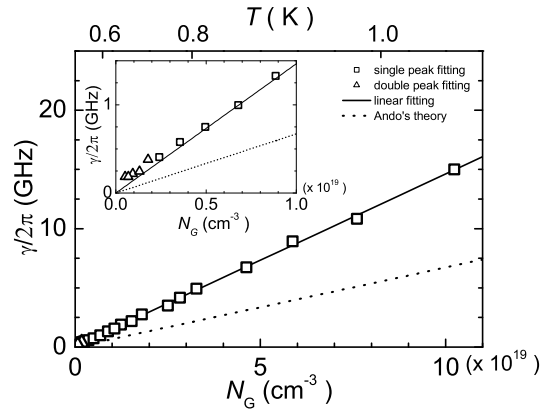}
\end{center}
\caption{\label{LinearFit_34}The SSE linewidth $\gamma$ as a function of $^3$He vapor density $N_{\textrm{G}}$.
The linewidths obtained by the single peak fitting are expressed as ($\blacksquare$) and the ones obtained by the double peak fitting are expressed as ($\triangle$).
The solid line shows a linear fitting to the experimental data, and the dotted line is a theoretical calculation by Ando~\cite{Ando78}.
The inset is an enlarged plot at low vapor densities.
}
\end{figure}

The exponential temperature dependence of $\gamma$ is clearly shown in Fig. ~\ref{LinearFit_34},
where we plot $\gamma$ as a function of $^3$He vapor density $N_{\textrm{G}}$. 
The linewidth is proportional to $N_{\textrm{G}}$, i.e. $\gamma = \alpha N_{\textrm{G}}$, where $\alpha$ is the proportionality coefficient, at $N_{\textrm{G}} \ge 1 \times 10^{18} {\mathrm {cm}}^{-3}$.
At low temperatures, $\gamma/(2 \pi)$ becomes constant at around 180~MHz.

We have found that the proportionality coefficient $\alpha$ derived from the least-square fitting of the linewidth data is larger than that of the Ando's calculation by a factor of 2.1. 
As is seen in Fig. \ref{LINEWIDTH}, the experimental linewidth on liquid $^4$He is also larger than the theoretical one by a factor of 1.6.
Therefore, the discrepancy in $\alpha$ is a common feature of both the $^3$He and $^4$He cases. 

At $T < 0.4 ~{\mathrm K}$, the resonance splits into 5 - 6 peaks. Interestingly, the total width, which is the frequency width from the transition tail of the leftmost peak to that of the rightmost one, is independent of temperature, from 0.4 K to the lowest temperature 10 mK. 
The total width is about 800 MHz, and the separation between neighboring peaks is typically 100~MHz.
In this temperature regime, the lineshape depends on the areal electron density: the total width decreases as $n_s$ decreases,
although the separations between the peaks are independent of $n_s$.

\section{Discussion}
\subsection{Linewidth at $T>0.4$~K}
The discrepancy in the linewidth of the theory~\cite{Ando78} and experiment is seen in both the case of $^3$He and $^4$He.
The linewidth data obtained by Grimes~\cite{Grimes74} on liquid $^4$He agrees well with our data, so they show the discrepancy to the theory as well.
In contrast, the data of Collin~\textit{et al.}~\cite{Collin02} shows an apparent agreement with Ando's theory. However, their data are highly scattered, by more than a factor of two.
Due to this inaccuracy, it is impossible to make a quantitative comparison with the theory. We therefore conclude that the discrepancy in the linewidth is substantial.
 
A similar discrepancy was found in the mobility measurement performed by Shirahama \textit{et al.}~\cite{Shirahama95}.
The mobility was smaller than the values calculated by Saitoh by a factor of 1.5 - 2~\cite{Saitoh77, Ando78}.
The discrepancies in the linewidth and the mobility may originate from a common mechanism,
and suggest that the theories underestimate the relaxation time due to electron - vapor atom scattering. 

\subsection{Inhomogeneous Broadening at Low Temperatures}
The present work has revealed that the intersubband transition linewidth is determined by electron - vapor atom scattering at modestly low temperatures ($0.4 < T < 1.1$~K). However, the multiple peak structure has prevented to study the contribution of electron - ripplon collisions.
Here we discuss some possible origins of the observed linewidth at low temperatures. 
In general, the linewidth is determined by two mechanisms: One is the intrinsic (homogeneous) broadening mechanism. In the SSE system, it is caused by the interactions of the SSE with surrounding environments, such as vapor atoms or ripplons. Another is the inhomogeneous broadening. The spatial inhomogeneity of the holding electric field $E_{\perp}$ is a possible source. This inhomogeneity can contribute to the linewidth.
The resonant frequencies of the intersubband transitions depend on $E_{\perp}$. If $E_{\perp}$ varies with the in-plane position of the electrons, so does the transition frequency. The spatial distribution of the transition frequency will broaden the linewidth.

Since the real SSE is a finite system, the electric field and electron density near the edge of the electron sheet are determined so as to satisfy proper boundary conditions and to minimize the energy of the system. Therefore, $E_{\perp}$ and $n_s$ are inevitably nonuniform near the edge. 
The parallelism between the electrodes and the electron sheet is also important to eliminate the inhomogeneous broadening. 
The tilt of the electrode surfaces from the horizontal causes a spatial gradient in the helium depth. This makes $E_{\perp}$ inhomogeneous: $E_{\perp}$ will be a smoothly varying function of the coordinates in the electron sheet. 

\begin{figure}[tbp]
\begin{center}
\includegraphics[width=7cm,keepaspectratio]{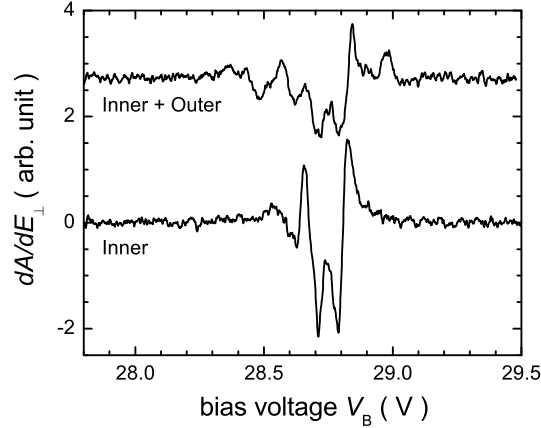}
\end{center}
\caption{\label{BOTH_INNER_15mK}
Two types of the lineshapes: The lower line (denoted as Inner) was measured by applying the pressing field modulation only to the inner Corbino electrode: i.e. the 
effect of the electrons near the sample edge is eliminated.  
The upper one (Inner $+$ Outer) was taken in a usual way. Both data were taken on liquid $^3$He at
$n_s=7.0\times10^7$~cm$^{-2}$ and $T$=0.01~K.}
\end{figure}

To investigate the inhomogeneous broadening effect, we have measured the absorption spectrum with applying the bias voltage modulation only to the inner part of the top Corbino electrode. By this way the derivative signal constitutes only from the electrons beneath the inner Corbino electrode, and the contribution from the near-edge electrons is removed. Figure~\ref{BOTH_INNER_15mK} shows a comparison of the spectrum. 
The total width certainly decreases to about two thirds by excluding the contribution from the near-edge electrons. 
We therefore conclude that the multiple peak structure is partly attributed to the field inhomogeneity at the edge, and also to the incomplete parallelism between the electron sheet and the electrodes. 
 
In the present work, the tilt of 1.5~milliradians might produce the linewidth of about 800~MHz. 
This is comparable to the total width of the multiple peaks seen below 0.4~K.
In order to obtain the proper lineshape at low temperatures where the electron - ripplon scattering is dominant, one needs to improve greatly the parallelism, 
within 0.01~milliradians, which decreases the linewidth down to a few MHz. 
Although such a high parallelism is in principle possible, it is not realistic to pursue the parallelism with the present setup: 
Employing much smaller electron sheets, and detecting locally or individually the intersubband transitions will be more promising.

The intersubband transitions could also be affected by the Coulomb interaction between the electrons.
The energy of an electron depends on the quantum state of its neighbor, because the distance between neighboring electrons increases with the quantum number.
This causes broadening of the linewidth.
However, the energy shift is estimated to be around 10~MHz, which is more than an order of magnitude smaller than the inhomogeneous broadening found in this experiment.
The Coulomb interaction is not enough to explain the observed broadening.

Collin \textit{et al.}~\cite{Collin02} derived the linewidth at low temperatures by the deconvolution of the observed lineshape, which has an inhomogeneous broadening of about 50~MHz.
We did not succeed to deconvolute the lineshape because of the asymmetry and multiple peak structure. For the precise determination of the linewidth in the ripplon scattering regime, some refinements of the experimental method are needed.

\section{Conclusions}
We have observed for the first time the intersubband transitions of SSE on $^3$He by directly measuring microwave absorption.
The absorption linewidth is proportional to the $^3$He vapor density above 0.4 K. 
This proportionality is well explained by the theory of Ando.
However, the experimental linewidth is larger than the theoretical result by a factor of two. 
This discrepancy is a common feature of the SSE both on $^3$He and on $^4$He: 
The theory may underestimate the electron - vapor atom scattering rate.
It is highly desirable to elucidate the origin of the discrepancy from the theoretical side.
 
At low temperatures where the vapor atom scattering ceases, the lineshape shows a multiple peak structure.
The structure is partly due to the spatial inhomogeneity in the static electric field.
It could be eliminated by improving the parallelism between the SSE and the electrodes, and employing smaller sample cells.
However, at the moment, it is not clear whether the other causes of the structure are also technical problems, or imply some new physical mechanisms. 
The detection of the intersubband transitions by a local probe, such as the SET, may solve this issue. 
 
\section*{Acknowledgment}
We thank H. Ikegami and Yu. P. Monarkha for helpful discussions. 
This work is supported by the Grant-in-Aid for Scientific Research from the Ministry of Education and Science in Japan.

\end{document}